\documentstyle [12pt] {article}
\begin{document}

\pagestyle{empty}
\rightline{\vbox{
\halign{&#\hfil\cr
&NUHEP-TH-93-10\cr
&April 1993\cr}}}
\bigskip
\bigskip
\bigskip
{\Large\bf
	\centerline{Electromagnetic Production of Quarkonium}
	\centerline{in $Z^{0}$ decay}}
\bigskip
\normalsize
\centerline{Sean Fleming}
\centerline{\sl Department of Physics and Astronomy, Northwestern University,
    Evanston, IL 60208}
\bigskip

\begin{abstract}
The decay $Z^{0}\rightarrow Q+ \ell^{+}\ell^{-}$, where $Q$ is a
$J^{PC}=1^{--}$ quarkonium state, has a very clean final state, which should
make it easy to detect. The branching ratio of this mode is greater than
$10^{-6}$ for $\rho$, $\phi$, and $\psi$, indicating that these processes may
be detectable at LEP.
\end{abstract}
\vfill\eject\pagestyle{plain}\setcounter{page}{1}
Well over one million $Z^{0}$ decay events have been accumulated at
LEP, and further improvements are expected. This makes it possible to
investigate rare decays of the $Z$-boson with branching ratios as small as
$10^{-6}$. Processes involving the production of the charmonium state $J/ \psi$
that have been considered previously are $\; Z^{0}\rightarrow\psi + gg$,
$\;Z^{0}\rightarrow\psi + c\bar{c}$, and $Z^{0}\rightarrow\psi + \gamma$
\cite{bcy,bck,gkpr,jm,k}. The production of a lepton pair and a $\psi$ from a
$Z^{0}$ is another example of such a process. This decay is especially
interesting because of its clean final state, and its relatively large
branching ratio. In this paper the rates for
$Z^{0}\rightarrow Q + \ell^{+} \ell^{-}$, where $Q$ is a $J^{PC}=1^{--}$
quarkonium state, are calculated to leading order in $\alpha$ in a model
independent way. The branching ratios for $\rho$, $\phi$, and $\psi$ are
greater than $10^{-6}$ which indicates that these processes may be
detectable at LEP.

The specific decay of $Z^{0} \rightarrow \psi \; + \; \ell^{+}\ell^{-}$ is
calculated first, and later generalized to other quarkonium states with
$J^{PC}=1^{--}$. The decay rate is
\begin{equation}
\Gamma(Z^{0}\rightarrow \psi + \ell^{+}\ell^{-})
=\frac{1}{2M_{Z}}\int[dk][dk'][dp] \; (2\pi)^{4}\delta^{4}
(Z-k-k'-p) \; \frac{1}{3}\sum |A|^{2},
\label{eq:lgo}
\end{equation}
where $k$, $\; k'$, $p$ and $Z$ are the 4-momenta of the $\; \ell^{+}$,
$\; \ell^{-}$, $\; \psi$, and $Z^{0}$, and $\; [dp]=d^{3}p/(16 \pi^{3}p_{0})$
is the Lorentz invariant phase space element. The two Feynman diagrams that
contribute to the amplitude $A$ at leading order in $\alpha$ are shown in
figure 1. Neglecting lepton mass, squaring the amplitude, averaging over
initial spins, and summing over final spins gives
\begin{eqnarray}
\frac{1}{3}\sum |A|^{2} & = &
\frac{16 \pi^{2}}{3} \; \frac{\alpha^{2} g^{2}_{w} g^{2}_{\psi}}
{\cos\theta_{w}} \; (C^{2}_{A}+C^{2}_{Q}) \; g(z_{1},z_{2}, \lambda)
\nonumber \\
g(z_{1},z_{2},\lambda) & = & \frac{1}{2}\left(\frac{z_{1}}{z_{2}}+
\frac{z_{2}}{z_{1}}\right) - \frac{\lambda}{2}\left(\frac{1}{z^{2}_{1}}+
\frac{1}{z^{2}_{2}}\right) + (1+\lambda) \! \left(\frac{1+\lambda}{z_{1}z_{2}}-
\frac{1}{z_{1}} - \frac{1}{z_{2}}\right),
\label{ampsq}
\end{eqnarray}
where $\lambda=M_{\psi}^{2}/M_{Z}^{2}$, $C_{A} = -1+4 \sin \theta_{w}$, and
$C_{V} = 1$. This is similar to the calculation done in Ref. \cite{hms} for the
$e^{+} e^{-} \rightarrow q \bar{q} g^{*}$ cross section for producing a virtual
gluon. Here $z_{1}=(Z-k)^{2}/M^{2}_{Z}$ and $z_{2}=(Z-k')^{2}/M^{2}_{Z}$,
$\; g_{w}=e/\sin\theta_{w}$ is the weak coupling constant, and $\theta_{w}$ is
the Weinberg angle. The strength of the $\psi$-photon coupling $g_{\psi}$ is
related to the matrix element of the electromagnetic current by
\begin{equation}
<\psi|\sum_{i}e_{i}q_{i}\gamma_{\alpha} \bar{q_{i}}|\not{\! 0}>=g_{\psi}
M_{\psi}^{2} \epsilon_{\alpha}(p).
\label{me}
\end{equation}
The coupling $g_{\psi}$ can be calculated from the rate for the quarkonium
state to decay to $e^{+} \; e^{-}$:
\begin{equation}
\Gamma_{e^{+}e^{-}}={4 \pi \over 3} \alpha^{2} M_{\psi} g^{2}_{\psi}.
\label{coup}
\end{equation}
Using (\ref{ampsq}) in (\ref{eq:lgo}) the rate reduces to
\begin{equation}
\Gamma(Z^{0}\rightarrow\psi \; \ell^{+}\ell^{-})= 4 \alpha^2
g^{2}_{\psi} \;  \Gamma(Z^{0}\rightarrow\ell^{+}\ell^{-})\int^{1}_{\lambda}
dz_{1}
\int^{1+\lambda -z_{1}}_{\frac{\lambda}{z_{1}}} dz_{2} \; g(z_{1},z_{2},
\lambda).
\label{rate}
\end{equation}
Integrating the function $g(z_{1},z_{2},\lambda)$ over $z_{2}$ and changing
variables from $z_{1}$ to $E_{\psi}$ yields the $\psi$ energy distribution
shown in figure 2. Note that the energy distribution in the figure does not
go to zero at $\sim \! M_{Z}/2$ because the lepton mass has been neglected in
the calculation. Including the lepton mass makes the curve go to zero at the
endpoint, but is negligible near the peak of the distribution. For $\psi$
energy $E_{\psi} \gg M_{\psi}$ the energy distribution has a simple analytic
form:
\begin{eqnarray}
\lefteqn{\frac{d \Gamma}{dE_{\psi}} (Z^{0}\rightarrow\psi(E_{\psi})+\ell^{+}
\ell{-}) =}
\nonumber \\
& & \frac{4 \alpha^{2} g^{2}_{\psi}}{M_{Z}}
\Gamma(Z^{0}\rightarrow\ell^{+}\ell^{-}) \left[
\frac{(z-1)^{2}+1}{z} \left( 4\log{z} \; - \; 2\log\lambda \right)
-4z \right]
\label{hel}
\end{eqnarray}
where $z=2 E_{\psi}/M_{Z}$. The coefficient of the logarithm is proportional to
the Altarelli-Parisi function for the splitting of a lepton into a photon of
momentum fraction $z$. Doing the remaining integral gives the full decay rate
\begin{eqnarray}
\lefteqn{ \Gamma(Z^{0}\rightarrow\psi \; \ell^{+}\ell^{-})= }
\nonumber \\
& & 4 \alpha^2g^{2}_{\psi} \;  \Gamma(Z^{0}\rightarrow\ell^{+}\ell^{-})
\left[ \frac{1}{2}(1+\lambda)^2\log^{2}\lambda \; + \;
\frac{1}{2}(3 + 4\lambda +3\lambda^{2})\log\lambda \; + \;  \frac{5}{2}(1 -
\lambda^{2}) \; - \; \right.
\nonumber \\
& & \left.
2(1+ \lambda)^{2}[Li_{2}(-\lambda) \; + \; \log(1 + \lambda)\log\lambda \;
+ \; \frac{\pi^{2}}{12} ] \right].
\label{intgfun}
\end{eqnarray}
where $Li_{2}(x)$ is the dilogarithm function. Neglecting the lepton mass
introduces an error of order $m^{2}_{\ell}/M^{2}_{Z}$ in the rate.

It is trivial to extend this calculation to other bound quarkonium states by
replacing $M_{\psi}$ by the quarkonium mass $M_{Q}$, and using the appropriate
leptonic decay rate $\Gamma_{e^{+}e^{-}}$ for the quarkonium to go into two
electrons. The energy distribution for the production of the $\rho$ and $\phi$
are also shown in figure 2. Note that all of the curves have peaks at
approximately $1.8 M_{Q}$. The rates for the decay of $Z^{0}$ into
$\ell^{+} \; \ell^{-}$ and $\; \rho$, $\; \phi$, $\; \psi$, $\; \psi '$,
$\; \Upsilon$, summed over all leptons, are given in table~1. The most visible
decay modes of the mesons are $\; \rho \rightarrow \pi^{+} \pi^{-}$ with a
branching ratio of $100 \% $, $\; \phi \rightarrow K^{+} K^{-}$ with a
branching ratio of $49 \%$, and
$\psi, \; \psi ', \; \Upsilon \; \rightarrow e^{+} e^{-} + \mu^{+}\mu^{-}$ with
branching ratios of  $\; 12.2 \%$, $\; 1.7 \%$, and $\; 5 \%$. This reduces the
number of events that can be seen at LEP to the number in the last column of
table 1. If a cutoff on the quarkonium energy of $E_{Q} = 10$ GeV is introduced
to account for the difficulties in detecting low energy particles the visible
events are further reduced by about $45 \%$, $50 \%$, $60 \%$ for $\rho$,
$\phi$, $\psi$.
\begin{table}
\begin{tabular}{|c|c|c|c|c|}
\hline
Onium State & Visible Decay & Branching Ratio    & Events per & Visible Events
\\
            & Products      & into Visible Modes & Million    & per Million
\\ \hline
$\rho$      & $\pi \pi$     & $100 \%$           & 27         & 27     \\
$\phi$      & $K^{+}K^{-}$  & $49 \%$            & 3.6        &  1.8   \\
$\psi$      & $e^{+}e^{-}+\mu^{+}\mu^{-}$ & $12.2 \%$ & 2.3   & 0.27   \\
$\psi '$    & $e^{+}e^{-}+\mu^{+}\mu^{-}$ & $1.7 \%$  & 0.66  & 0.011  \\
$\Upsilon$  & $e^{+}e^{-}+\mu^{+}\mu^{-}$ & $5 \%$    & 0.06  & 0.003  \\
\hline
\end{tabular}
\caption{Events and visible events per million $Z^{0}$ decays for
$Z^{0}\rightarrow \ell^{+}\ell^{-}+ Onium$, summed over $\ell=e,\mu,\tau$.}
\end{table}
The branching ratio for $\; Z^{0}\rightarrow \psi + \ell^{+}\ell^{-}$, summed
over $\; \ell=e,\mu,\tau \; $ is $\; 2.3 \times 10^{-6}$ while the branching
ratios for other $\psi$ production mechanisms are:
B$(\; Z^{0}\rightarrow\psi +c\bar{c})=2.2 \times 10^{-5}$,
B$(\; Z^{0}\rightarrow\psi +\gamma)$\footnote{A smaller rate has been reported
\cite{jm}} = $5.46 \times 10^{-8}$, and
B$(\; Z^{0}\rightarrow\psi +gg)=2.18\times 10^{-7}$ \cite{bcy,gkpr,k}. Only
the rate for $\;Z^{0}\rightarrow\psi + c\bar{c}\;$ is greater than the rate
calculated in this paper. Despite this the process
$\; Z^{0}\rightarrow \psi + \ell^{+}\ell^{-}$ may be more detectable since
there is a large background to $\;Z^{0}\rightarrow\psi + c\bar{c}$ due to the
decay $B\rightarrow \psi$ \cite{bck}. Also note that the rate for the decay
$Z^{0}\rightarrow \psi + \ell^{+} \ell^{-}$ is an order of magnitude greater
than the rate for $Z^{0}\rightarrow \psi + \gamma$, even though the former
process is of order $\alpha^{2}$ while the latter process is of order $\alpha$.
The reason for this is that $Z^{0}\rightarrow\psi +\gamma$ is a short distance
process with the $c$ and $\bar{c}$ that form the $\psi$ being produced in a
region with size of order $1/M_{Z}$, which suppresses the rate by a factor of
$1/M^{2}_{Z}$. In the process $\; Z^{0}\rightarrow \psi + \ell^{+}\ell^{-}$
the dominant mechanism for $\psi$ production is fragmentation of a lepton or a
photon. Fragmentation is the process where a high energy parton splits into a
collinear $\psi$ with $E \gg M_{\psi}$ and a parton. In the fragmentation
process, the $c$ and $\bar{c}$ that form the $\psi$ are produced in a region
with size of order $1/m_{c}$, so that the rate is only suppressed by a factor
of $1/m^{2}_{c}$ \cite{flem}. Fragmentation is also responsible for the rate
for $ \; Z^{0}\rightarrow\psi +c\bar{c}$ being two orders of magnitude larger
than the rate for $Z^{0}\rightarrow\psi +gg$, in spite of the fact that both
rates are the same order in $\alpha_{s}$. The process
$Z^{0}\rightarrow\psi +gg$ is a short distance process that is suppressed by a
factor of $m^{2}_{c}/M^{2}_{Z}$, while
$ \; Z^{0}\rightarrow\psi+c\bar{c}$ includes a fragmentation contribution that
has no such suppression \cite{bcy}.

With over a million $Z^{0}$ decay events collected at LEP an investigation of
rare modes is a realistic prospect. The production of the $\psi$ is
particularly interesting since the heavy charmed quarks involved allow
calculations to be done using perturbative quantum chromodynamics (QCD).
The analysis in Ref. \cite{bcy} of $ \; Z^{0}\rightarrow\psi +c\bar{c}$ makes
it clear that fragmentation plays an important role in $\psi$ production.
Unfortunately there will be a large background to this process from $B$ meson
decay. The decay $Z^{0}\rightarrow\psi+\ell^{+}\ell^{-}$ is appealing because
of the fragmentation contribution, and its particularly clean final state. The
rate is large compared to $Z^{0}\rightarrow\psi+ \gamma$, and may even be
detectable at LEP. The rate for the production of the lighter quarkonium
states $\rho$ and $\phi$ are even larger and should certainly be detectable
at LEP.

This work is supported in part by the U.S. Department of Energy, Division of
High Energy Physics, under Grant DE-FG02-91-ER40684. I wish to thank E. Braaten
for many helpful discussions.

\vfill\eject

\newpage
\noindent{\Large\bf Figure Captions}
\begin{enumerate}
\item The two Feynman diagrams for $Z^{0} \rightarrow \psi \; + \;
         \ell^{+} \ell^{-}$
        at leading order in $\alpha$.
\item The energy distribution of the decay rate for $\rho$, $\phi$ and $\psi$
       production.

\end{enumerate}

\vfill\eject

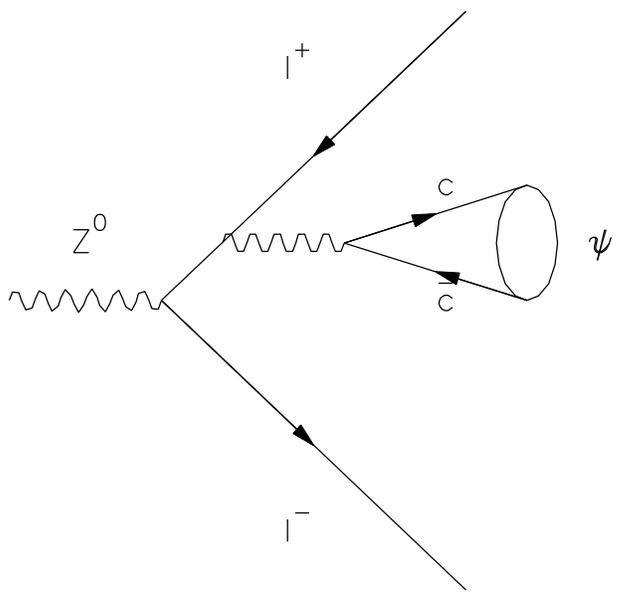
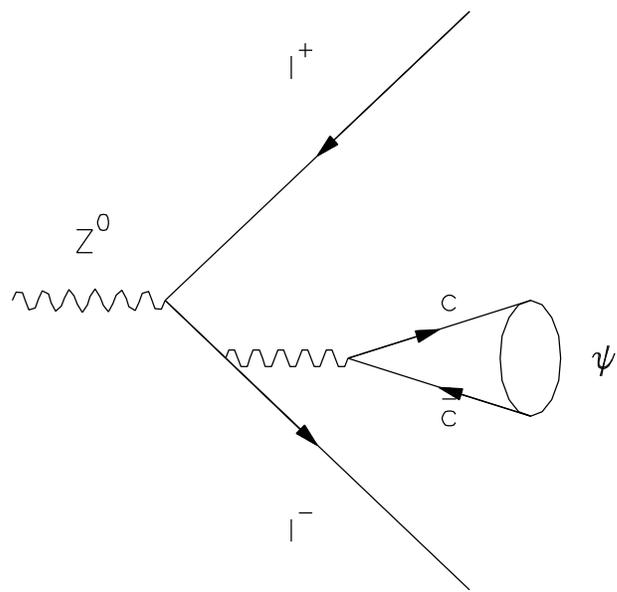
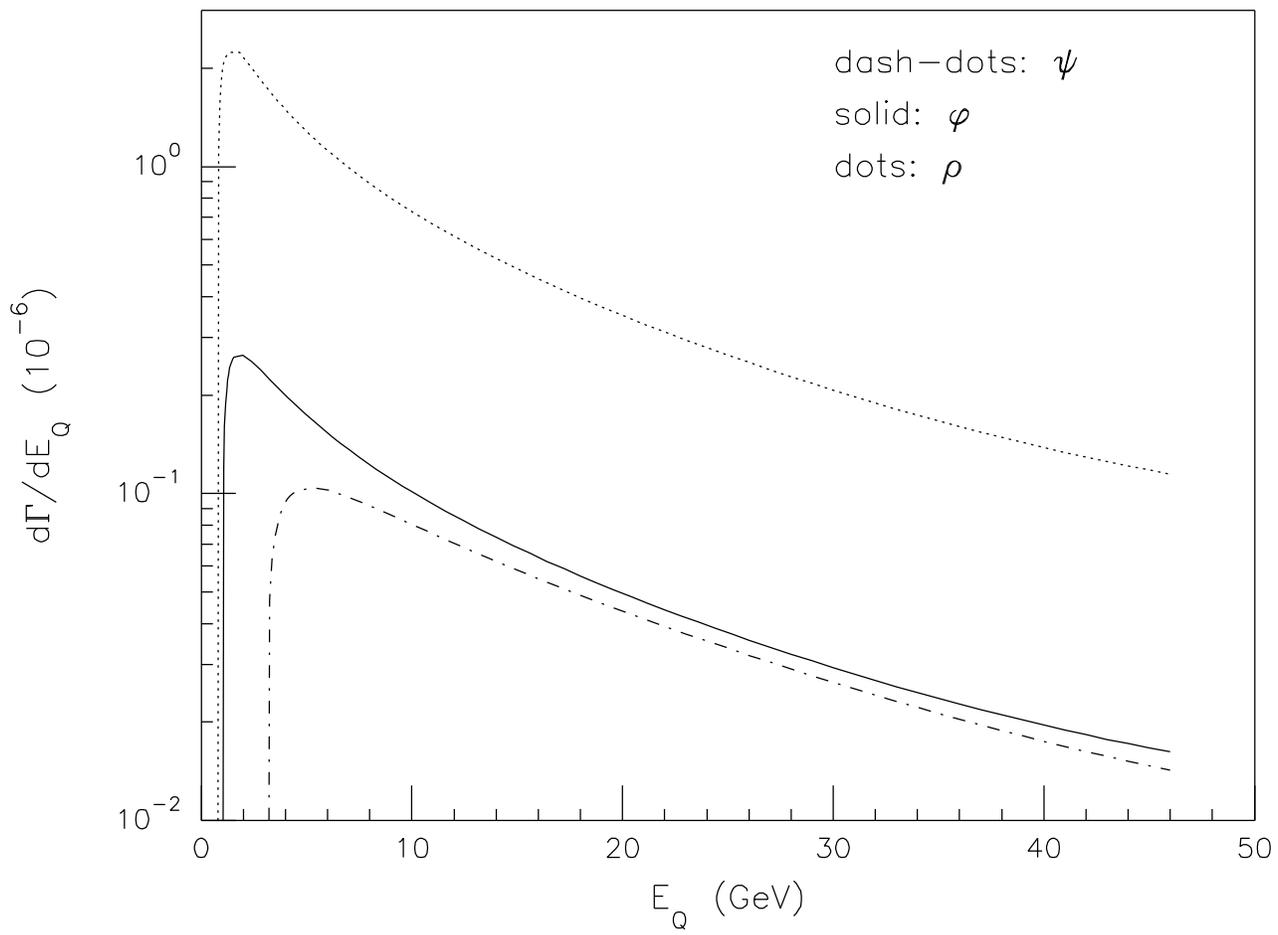
\end{document}